\documentclass[12pt]{iopart}
\usepackage{iopams}
\usepackage{setstack}
\usepackage{graphicx}
\usepackage{color}
\usepackage{cases}

\begin{document}

\title{A self-consistent theory of localization in nonlinear random media}

\author{Nicolas Cherroret}

\address{Laboratoire Kastler Brossel, UPMC-Sorbonne Universit\'es, CNRS, ENS-PSL Research University, Coll\`{e}ge de France, 4 Place Jussieu, 75005 Paris, France}
\ead{cherroret@lkb.upmc.fr}
\vspace{10pt}

\begin{abstract}
The self-consistent theory of localization is generalized to account for a weak quadratic nonlinear potential in the wave equation. For spreading wave packets, the theory predicts the destruction of Anderson localization by the nonlinearity and its replacement by algebraic subdiffusion, while classical diffusion remains unaffected. In 3D, this leads to the emergence of a subdiffusion-diffusion transition in place of the Anderson transition. The accuracy and the limitations of the theory are discussed.
\end{abstract}

%
%
%
%
%

\section{Introduction}

Anderson localization is a striking manifestation of wave interference in disordered systems \cite{Anderson58}. It has recently stirred considerable experimental interest, witness experiments on acoustic waves \cite{Hu08, Faez09}, microwaves \cite{Chabanov00, Chabanov01} and light \cite{Storzer06, Maret12, Schwartz07, Lahini09, Segev13} in disordered media, and on cold atomic matter waves in random potentials \cite{Billy08, Roati08, Jendrzejewski12, Kondov11, Semeghini15, Chabe08}. In this latter line of experiments, the treatment of localization is complicated by interactions between atoms. For weakly interacting bosons at low temperatures, the problem can be tackled within the single-particle formalism of the Gross-Pitaevskii equation \cite{Pitaevskii03}. In this description, atoms that evolve in the random potential additionally experience a quadratic nonlinear potential proportional to the atomic density. 
In optics, a similar formalism emerges from the scenario of paraxial light  propagation in disordered media displaying a Kerr nonlinearity \cite{Schwartz07, Lahini09, Lahini08, Pertsch04}. 

The effect of a weak nonlinear potential proportional to the density on Anderson localization has been studied in a number of theoretical works dealing with stationary solutions of the nonlinear wave equation \cite{Iomin07, Fishman08, Bodyfelt10}, transport of stationary flows through a disordered region \cite{Gredeskul92, Paul05, Paul07, Hartung08, Wellens08, Wellens09a, Wellens09b}, thermalization \cite{Mulansky09, Basko11, Cherroret15}, speckle instability \cite{Skipetrov00, Bortolozzo11} or wave packet spreading \cite{Kopidakis08, Pikovsky08, Flach09, Iomin09, Soffer08, Wang09, GarciaMata09, Flach10, Iomin10, Veksler09, Laptyeva10, Cherroret14}. These various problems
are not always directly connected because the superposition principle no longer holds in a presence of nonlinear effects. This makes the interplay between Anderson localization and nonlinearity a multiform and very rich question.  

In this paper, we build the foundations of a self-consistent theory (SCT) of Anderson localization in situations where a quadratic nonlinear potential is present in the wave equation, using the Gross-Pitaevskii equation as a paradigmatic example. This theory is a generalization of the SCT of localization invented in the 80s \cite{Vollhardt80, Vollhardt82, Wolfle92, Vollhardt10}, and relies on recent ideas developed in the context of nonlinear coherent backscattering \cite{Agranovich91, Hartung08, Wellens08, Wellens09a, Wellens09b, Hartmann12} and of Anderson localization in the stationary limit of nonlinear wave propagation \cite{Eckert10}. We apply this novel framework to the scenario where a wave packet spreads in a random potential in three dimensions (3D), for which we find that a weak nonlinearity destroys Anderson localization and gives rise to a subdiffusive motion below the Anderson critical point. To leading order, the diffusive regime above the critical point and the critical point itself remain, on the other hand, qualitatively not affected by the nonlinearity. This suggests the emergence of a nonlinearity-driven subdiffusion-diffusion transition, at least at intermediate times. In agreement with previous work \cite{Kopidakis08, Pikovsky08, Flach09}, in one dimension (1D) the nonlinear SCT also predicts a subdiffusive motion at long times, although with a slightly overestimated subdiffusive exponent.

The paper is organized as follows. In Sec. \ref{framework}, the general problem of expansion of a wave packet in a random potential is formulated, starting from the single-particle Schr\"odinger equation. From this formalism, the essential lines of the SCT of localization and his physical predictions are recalled in Sec. \ref{Linear_SCT_Sec}. An extension of the SCT based on the Gross-Pitaevskii equation is then presented in Sec. \ref{Nonlinear_SCT_sec}, and solved by an approximate method in Sec. \ref{Approx_sol_Sec}. In particular, the question of the effect of the nonlinearity on the Anderson transition is addressed. Sec. \ref{TPST} is devoted to the generalization of the one-parameter scaling theory of localization to the nonlinear regime. The accuracy and the limitations of our approach are finally discussed in Sec. \ref{Validity_Sec}.

\section{Framework}
\label{framework}

\subsection{Average density}

In this section, we leave out the nonlinearity and discuss the linear time evolution of a wave packet $\psi(\bi{r},t)$ in a random potential $V(\bi{r})$, with the initial condition $\psi(\bi{r},t=0)\equiv\phi(\bi{r})$. The Fourier component $\psi_\epsilon(\bi{r})=\int_{-\infty}^{\infty} dt\, e^{i\epsilon t}\psi(\bi{r},t)$ obeys the stationary Schr\"odinger equation
\begin{equation}
\label{Schro_eq_eps}
\left[-\epsilon-\frac{1}{2m}\boldsymbol\nabla^2+V(\bi{r})\right]\psi_\epsilon(\bi{r})=0,
\end{equation}
where we have set $\hbar=1$. In the following, we assume for simplicity that $V(\bi{r})$ is Gaussian distributed, and uncorrelated \cite{Akkermans07}:
\begin{equation}
\label{WN}
\overline{\delta V(\bi{r})\delta V(\bi{r}')}=
\gamma\delta(\bi{r}-\bi{r}'),
\end{equation}
where $\gamma=1/(2\pi\rho\tau)$, with $\tau$ the mean free time and $\rho$ the density of states per unit volume.
The figure of merit of this paper is the disorder-averaged density $n(\bi{r},t)\equiv\overline{|\psi(\bi{r},t)|^2}$.
To calculate this quantity, we introduce the retarded (advanced) Green's function $G^R_\epsilon(\bi{r}',\bi{r})$ [$G^A_\epsilon(\bi{r}',\bi{r}$)] of Eq. (\ref{Schro_eq_eps}), so that
\begin{equation}
\label{psi_G}
\psi_\epsilon(\bi{r})=\int d^d\bi{r}'G^R_\epsilon(\bi{r}',\bi{r})\phi(\bi{r}'),
\end{equation}
where $d=1,2,3$ is the dimensionality. Using Eq. (\ref{psi_G}), we find after a straightforward calculation \cite{Shapiro12}
\begin{equation}
\label{n_general}
n(\bi{r},t)=
\int_{-\infty}^{\infty} \frac{d\epsilon}{2\pi}
\int d^d\bi{r}'\int \frac{d^d\bi{k}}{(2\pi)^d}
P_\epsilon(\bi{r}',\bi{r},t)
A_\epsilon(\bi{k})
W(\bi{r}',\bi{k},t=0).
\end{equation}
The various quantities appearing in Eq. (\ref{n_general}) are defined as follows. $A_{\epsilon}(\bi{k})=-2\Im[\overline{G}^R_\epsilon(\bi{k})]$ is the spectral function, i.e. the energy distribution of a quantum particle with momentum $\bi{k}$. $P_\epsilon(\bi{r}',\bi{r},t)=\overline{G^R_\epsilon(\bi{r}',\bi{r})G^A_\epsilon(\bi{r}',\bi{r})}/(2\pi\rho)$ is the propagator in the random potential, namely the probability density for a particle with energy $\epsilon$ to propagate from point $\bi{r}'$ to point $\bi{r}$ \cite{Akkermans07}. Finally, $W(\bi{r}',\bi{k},t=0)\equiv\int d^d\brho\,e^{-i\bi{k}\cdot\brho}\phi(\bi{r}'+\brho/2)\phi^*(\bi{r}'-\brho/2)$ is the Wigner distribution of the initial wave packet. 

Eq. (\ref{n_general}) has the following interpretation: when released in a random potential, a particle of momentum $\bi{k}$, initially located at point $\bi{r}'$, acquires an energy $\epsilon$ with probability $A_\epsilon(\bi{k})$ and propagates over a time span $t$ with this energy up to point $\bi{r}$ with probability $P_\epsilon(\bi{r}',\bi{r},t)$. The density at $\bi{r}$ then follows from integration over all initial positions, momenta and energies.

\subsection{Quasi-monochromatic wave packet}
\label{MonoC}

A wave packet prepared in a random potential has a more or less broad momentum distribution. 
For this reason, its global dynamics may be difficult to trace because it results from a complicated superposition of many energy components $\epsilon$, see Eq. (\ref{n_general}) \cite{Muller15}. 
To circumvent this issue, in this paper we restrict ourselves to the ``quasi-monochromatic'' limit: we first assume that the momentum width of the Wigner function is small compared to that of the spectral function, such that $\int d^d\bi{k}/(2\pi)^dA_\epsilon(\bi{k})W(\bi{r}',\bi{k},t=0)\simeq A_\epsilon(\bi{k}_0) |\phi(\bi{r}')|^2$, where $\bi{k}_0$ is the mean momentum of the initial packet. Second, we make use of an on-shell approximation for the energy distribution: $A_\epsilon(\bi{k}_0)\simeq 2\pi\delta(\epsilon-\epsilon_0)$, with $\epsilon_0=\bi{k}_0^2/(2m)$, so Eq. (\ref{n_general}) reduces to
\begin{equation}
\label{n_simple_2}
n(\bi{r},t)\simeq
\int d^d\bi{r}'\,
P_{\epsilon_0}(\bi{r}',\bi{r},t)
|\phi(\bi{r}')|^2.
\end{equation}
Possible issues raised by the quasi-monochromatic limit will be discussed in Sec. \ref{Validity_Sec}. For an initially localized wave packet $|\phi(\bi{r}')|^2\sim\delta(\bi{r}'-\bi{r}_0)$, Eq. (\ref{n_simple_2}) implies that the dynamics of $n(\bi{r},t)$ is directly governed by $P_{\epsilon_0}(\bi{r}',\bi{r}_0,t)$, which is the object we discuss in the next section. In the following, to lighten the notations we will rename  the central energy (momentum) $\epsilon_0$ ($k_0$) of the wave packet to $\epsilon$ ($k$), and we will drop the energy index in the propagator, writing $P$ instead of $P_\epsilon$.

\section{Self-consistent theory in linear random media}
\label{Linear_SCT_Sec}

\subsection{Ladder diagrams and diffusion}

The propagator $P\equiv\overline{G^RG^A}/(2\pi\rho)$ obeys the Bethe-Salpeter equation \cite{Akkermans07}
\begin{equation}
\label{BSE}
\overline{G^RG^A}=
[\overline{G}^R\overline{G}^A]+[\overline{G}^R\overline{G}^A]\, R_\mathrm{tot}\, [\overline{G}^R\overline{G}^A].
\end{equation}
The ballistic component $\overline{G}^R\overline{G}^A$ decays exponentially on the scale of the mean free path and is thus negligible at large distances \cite{Akkermans07}. The kernel $R_\mathrm{tot}$ contains all contributions involving multiple scattering. When interference can be neglected on average,
it reduces to the so-called series of ladder diagrams $R$, which is solution of the iterative equation
\begin{figure}[h]
\centering
\includegraphics[width=11.5cm]{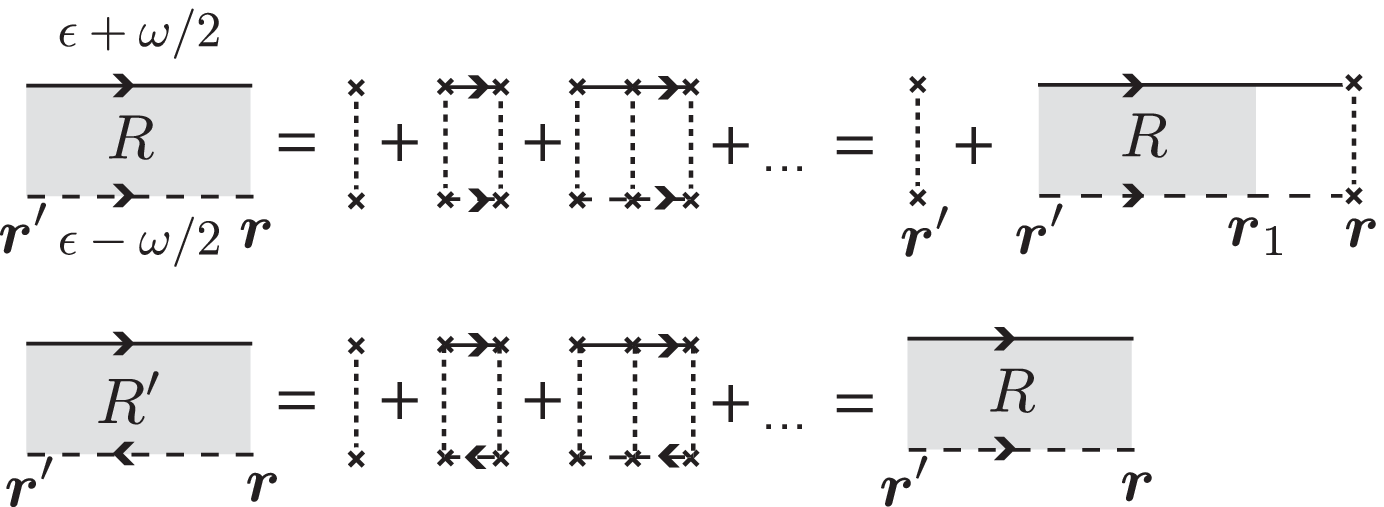}
\caption{\label{LCLC} 
Top: series of ladder diagrams $R$. Solid lines refer to the retarded Green's function $G^R$ and dashed lines to the advanced Green's function $G^A$. Dotted lines symbolize the correlation function (\ref{WN}) and arrows indicate the direction of propagation. Bottom: series of crossed diagrams $R'$, obtained by time-reversing one amplitude in the ladder sequence. Due to time-reversal invariance, $R'=R$.
}  
\end{figure}
\begin{equation}
\label{RL_sequence}
\fl R(\bi{r}',\bi{r},\omega)=\gamma\delta(\bi{r}-\bi{r}')+
\gamma \int d^d\bi{r}_1 \overline{G}^R_{\epsilon+\omega/2}(\bi{r}_1,\bi{r})\overline{G}^A_{\epsilon-\omega/2}(\bi{r}_1,\bi{r}) R(\bi{r}',\bi{r}_1,\omega).
\end{equation}
The diagrammatic representation of Eq. (\ref{RL_sequence}) is shown in the top panel of Fig. \ref{LCLC}. 
In the hydrodynamic limit of long times $t\gg\tau$ and large distances $|\bi{r}-\bi{r}'|\gg\ell$ (where $\ell=k\tau/m$ is the mean free path),
Eqs. (\ref{BSE}) and (\ref{RL_sequence}) lead to $P(\bi{r}',\bi{r},\omega)=(\tau/\gamma) R(\bi{r}',\bi{r},\omega)$, where $P$ obeys the diffusion equation
\begin{equation}
\label{diff_eq}
\left(-i\omega-D_0\boldsymbol\nabla^2_{\bi{r}}\right)P(\bi{r}',\bi{r},\omega)=\delta(\bi{r}-\bi{r}'),
\end{equation}
with $D_0=k\ell/(dm)$ the diffusion coefficient. Eq. (\ref{diff_eq}) is conveniently solved in Fourier space, yielding $P(\bi{q},\omega)=1/(-i\omega+D_0\bi{q}^2)$, a quantity known as the \emph{diffuson}. 

\subsection{Weak localization}
\label{lin_WL}

In the diffusive description (\ref{diff_eq}), interference effects are discarded on average. In 3D, this is a good approximation when $k\ell\gg1$. Leading-order interference corrections to Eq. (\ref{diff_eq}) are known as weak localization \cite{Lee85}. Their account is based on the idea that when $k\ell$ is decreased, the probability for a particle to return to a region already visited increases, allowing for two amplitudes to ``swap'', as depicted by the diagram in the left panel of Fig. \ref{Loops_on_loops}. This mechanism involves a ``loop'' made of two counter-propagating scattering paths.
These two paths form a sequence given by the so-called series of crossed diagrams $R'$, to which we associate a propagator $P'=(\tau/\gamma)R'$ called the \emph{Cooperon}. The kernel $R'$ is represented in the bottom panel of Fig. \ref{LCLC}. For a time-reversal invariant system, $R'$ is equal to $R$ by virtue of the reciprocity principle \cite{Landau77}. The inclusion of  localization loops in the Bethe-Salpeter equation (\ref{BSE}) leads again to a diffusion equation for $P$, except that the diffusion coefficient $D_0$ is now modified by a weak localization correction proportional to the ``return probability'' $P'(\bi{r},\bi{r},\omega)=P(\bi{r},\bi{r},\omega)\equiv\int d^d\bi{Q}/(2\pi)^d/(-i\omega+D_0\bi{Q}^2)$ \cite{Wolfle92}:
\begin{subnumcases}{}
\displaystyle{P(\bi{q},\omega)=\frac{1}{-i\omega+D(\omega)\bi{q}^2}},\ \ \mathrm{with}\label{WL_a}  \\
       \displaystyle{\frac{1}{D(\omega)}=\frac{1}{D_0}+\frac{1}{\pi\rho D_0}\int\frac{d^d\bi{Q}}{(2\pi)^d}\frac{1}{-i\omega+D_0\bi{Q}^2}}.  \label{WL_b} 
\end{subnumcases}{}
Let us briefly comment on these formulas. First, Eq. (\ref{WL_b}) is the result of a perturbation theory. In 3D for instance, this implies that it holds only in the weak-disorder regime $k\ell\gg1$, where it predicts $|D(\omega\to 0)-D_0|/D_0\sim 1/(k\ell)^2\ll 1$.
Second, the interference correction naturally introduces a dependence of $D$ on $\omega$, which indicates that weak localization has a certain dynamics. In 3D, this dynamics takes place at the scale of the mean free time and is thus very fast. As shown below, this is no longer the case at the onset of Anderson localization.

\subsection{Anderson localization: self-consistent theory}
\label{linear_SCT}

\begin{figure}[h]
\centering
\includegraphics[width=9cm]{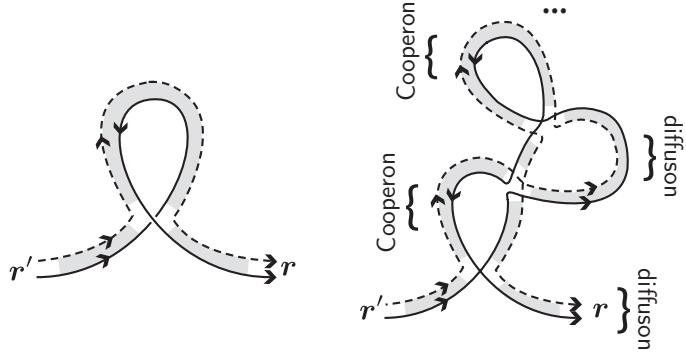}
\caption{\label{Loops_on_loops} 
Left: diagrammatic representation of weak localization: the diffusion coefficient is reduced due to the occurrence of loops in the propagation. Right: description of Anderson localization in the self-consistent scheme: loops are nested according to a diffuson/Cooperon alternation.
}  
\end{figure}

A prescription to go beyond the perturbative treatment and thus to describe Anderson localization was proposed by Vollhardt and W\"olfle in 1980 and is today known as the self-consistent theory (SCT) of localization \cite{Vollhardt80, Wolfle92, Vollhardt10}. The strategy consists in substituting $D(\omega)$ for $D_0$ in the return probability itself. As illustrated in the right panel of Fig. \ref{Loops_on_loops}, this procedure amounts to nesting localization loops, which allows for a cumulative treatment of interference.
It leads to \cite{Wolfle92}
\begin{subnumcases}{}
\displaystyle{P(\bi{q},\omega)=\frac{1}{-i\omega+D(\omega)\bi{q}^2}} \label{SCTa}\\
        \displaystyle{\frac{1}{D(\omega)}=\frac{1}{D_0}+\frac{1}{\pi\rho D_0}\int\frac{d^d\bi{Q}}{(2\pi)^d}\frac{1}{-i\omega+D(\omega)\bi{Q}^2}} \label{SCTb}.
\end{subnumcases}{}
As the SCT only takes into account certain classes of diagrams and neglects many others, it only provides an \emph{approximate} description of Anderson localization. Despite this, it rather accurately predicts the phase diagram of the Anderson model \cite{Kroha90}. It also fairly well  describes the dynamics of localization \cite{Lobkis05} and provides a quantitative basis for the phenomenological scaling theory \cite{Vollhardt82}, which are the two main problems we are interested in in this paper. On the other hand, the SCT fails at describing the large fluctuations in the vicinity of the critical point in 3D, and in particular gives a wrong estimation of critical exponents (see below).

\subsection{Predictions of the self-consistent theory}
\label{SCT_predictions}

To calculate the diffusion coefficient, we perform the $\bi{Q}$ integral in Eq. (\ref{SCTb}). For $d=2,3$, the latter displays an ultraviolet divergence, which signals the failure of the theory at short scales. Indeed, the SCT is  an \emph{hydrodynamic} theory and is thus designed to describe only large distances $|\bi{r}-\bi{r}'|\gg\ell$ and long times $t\gg\tau$, or equivalently $|\bi{q}|\ell\ll1$ and $\omega\tau\ll1$. This divergence is regularized by a cutoff $Q_\mathrm{max}\propto \ell^{-1}$. The prefactor is some extent arbitrary at our level of approximation (see however Sec. \ref{exp_sec}). Here, we make the usual choices $Q_\mathrm{max}=\ell^{-1}$ in 2D, which sets the prefactor of the localization length to unity, and $Q_\mathrm{max}=\pi/(3\ell)$ in 3D, which locates the position of the critical point at $k\ell=1$ (Ioffe-Regel criterion \cite{Ioffe60}, see below). The predictions of Eq. (\ref{SCTb}) for $D(\omega\to 0)$ are summarized in Table \ref{table1} (for more details of this calculation, we refer the reader to \cite{Shapiro82, Houches09}).
\begin{table}[h]
\centering
\begin{tabular}{|c|c|c|c|}
   \hline
    \raisebox{0.5ex}{$d=1$}& \multicolumn{3}{|c|}{ \raisebox{0.5ex}{$-i\omega\xi^2$,  $\mathrm{with}\ \xi=2\ell$}} \\
   \hline
   \raisebox{0.5ex}{$d=2$} &\multicolumn{3}{|c|}{ \raisebox{0.5ex}{$-i\omega\xi^2$,  $\mathrm{with}\ \xi=\ell\sqrt{\exp(\pi k\ell)-1}$}} \\
   \hline
   & \raisebox{0.5ex}{$k\ell>1$}& \raisebox{0.5ex}{$k\ell=1$}& \raisebox{0.5ex}{$k\ell<1$}\\
   \raisebox{4ex}{$d=3$} 
   	& \raisebox{1.5ex}{$\displaystyle{D_0\left[1-\frac{1}{(k\ell)^2}\right]}$} 
   		&  \raisebox{1.5ex}{$\displaystyle{\left(\frac{3D_0\ell}{2}\right)^{2/3}(-i\omega)^{1/3}}$}
			&  \raisebox{1.5ex}{$-i\omega\xi^2$,  $\,\displaystyle{\xi\simeq\frac{3\ell}{4(1-k\ell)}}$ for $k\ell\to 1^-$} \\
   \hline
\end{tabular}
\caption{\label{table1} Solution $D(\omega)$ of Eq. (\ref{SCTb}) in the limit $\omega\to 0$.}
\end{table}
This asymptotic limit is interesting because it is directly related to the long-time dependence of the mean square width $\langle \bi{r}^2(t)\rangle$ of the propagator $P(\bi{r}',\bi{r},t)$ that governs the wave-packet dynamics, according to
\begin{equation}
\label{D_width}
\langle \bi{r}^2(t)\rangle\underset{t\to\infty}{\propto} tD(i/t).
\end{equation}
In 1D and 2D, $D(\omega)=-i\omega\xi^2$ for any value of $k\ell$, such that $P(\bi{r}',\bi{r},t)\sim\exp(-|\bi{r}'-\bi{r}|/\xi)$ [$\langle \bi{r}^2(t)\rangle\propto\xi^2$], a clear hallmark of Anderson localization. In 3D, the asymptotic behavior of $D(\omega)$ depends on $k\ell$. For $k\ell>1$, $D(\omega\to 0)\equiv D=$ constant, so that $P(\bi{r}',\bi{r},t)=\exp[-|\bi{r}'-\bi{r}|^2/(4Dt)]/(4\pi D t)^{3/2}$ reduces to a diffusion kernel [$\langle  \bi{r}^2(t)\rangle\propto Dt$]. 
For $k\ell<1$ on the other hand, $D(\omega\rightarrow0)=-i\omega\xi^2$, which again signals Anderson localization. Exactly at $k\ell=1$ finally, $D(\omega\to 0)\propto(-i\omega)^{1/3}$ and $P(\bi{r}',\bi{r},t)$ has a subdiffusive character: $\langle  \bi{r}^2(t)\rangle\propto t^{2/3}$. The fact that $D\to0$ for $k\ell\to 1^+$ (and $\xi\to\infty$ for $k\ell\to 1^-$) indicates that $k\ell=1$ is the critical point of the Anderson transition. 

\begin{figure}[h]
\centering
\includegraphics[width=9cm]{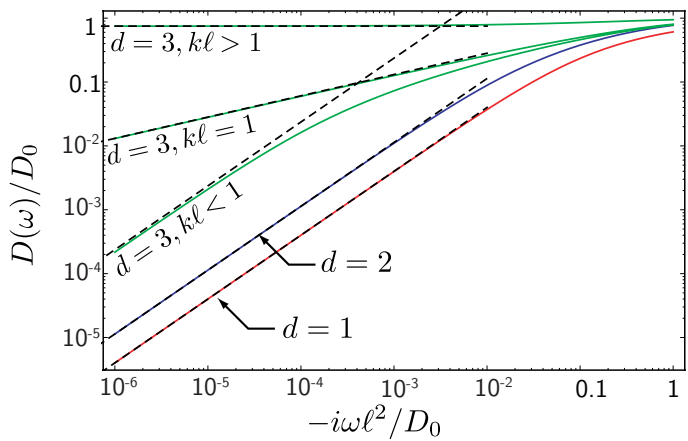}
\caption{\label{Linear_SCT} 
Solution $D(\omega)$ of Eq. (\ref{SCTb}) in 1D, in 2D for $k\ell=0.8$ and in 3D for $k\ell=0.95$, 1 and 2 (solid curves). Dashed lines are the asymptotic limits $D(\omega\to0)$ shown in Table \ref{table1}.
}  
\end{figure}

For  completeness, we show in Fig. \ref{Linear_SCT} the full frequency dependence of $D(\omega)$ predicted by Eq. (\ref{SCTb}) for $d=1$, $d=2$, and $d=3$ for three values of $k\ell$. The asymptotic results in Table \ref{table1} are shown as dashed lines. 

\subsection{Experimental considerations}
\label{exp_sec}

The SCT has proven to successfully describe experiments on wave localization, in particular in the context of transport of classical waves in 3D \cite{Hu08} and 2D \cite{Riboli16}. In practice, the theory requires as input the wave number $k$, the mean free path $\ell$ and the diffusion coefficient $D_0$. The ultraviolet cutoff $Q_\mathrm{max}=a/\ell$ is usually expressed in terms of an unknown numerical prefactor $a$ taken as a fit parameter. Note that while in this paper we restrict ourselves to transport in unbounded systems (a situation that can be realized in cold-atomic setups), the SCT can be readily generalized to describe open media, more commonly used in experiments on classical waves. In this case, translation invariance is lost, so the return probability $P'(\bi{r},\bi{r},\omega)$ -- and consequently  the diffusion coefficient $D(\bi{r},\omega)$ -- becomes explicitly position dependent \cite{Tiggelen00, Skipetrov06, Cherroret08}. An analogous extension can be performed in the nonlinear regime discussed hereafter.

\section{Self-consistent theory in nonlinear random media}
\label{Nonlinear_SCT_sec}

\subsection{Framework}
\label{Framework_NL}

We now address the nonlinear problem
\begin{equation}
\label{GPE}
\left[\frac{-1}{2m}\boldsymbol\nabla^2+V(\bi{r})+g|\psi(\bi{r},t)|^2\right]\psi(\bi{r},t)=i\frac{\partial\psi(\bi{r},t)}{\partial t},
\end{equation}
where $\int d^d\bi{r}|\psi(\bi{r},t)|^2=N$. In the cold-atomic context, Eq. (\ref{GPE}) is known as the Gross-Pitaevskii equation and governs the motion of a weakly interacting Bose-Einstein condensate of $N$ atoms \cite{Pitaevskii03}. 
We wish to analyze how the nonlinear term in Eq. (\ref{GPE}) affects the SCT of localization, to leading order in the (small) parameter $g$. A diagrammatic analysis of the stationary version of this problem [$\psi(\bi{r},t)=\psi(\bi{r})e^{i\epsilon t}$] was accomplished in  \cite{Wellens08, Wellens09a, Wellens09b}
in the limit $k\ell\gg 1$, and later extended to the treatment of strong localization in \cite{Eckert10}. In the following, we adapt these approaches to the non-stationary scenario.

\subsection{Ladder and crossed diagrams}
\label{LCD_nonlinear_sec}

The effect of the nonlinear term in Eq. (\ref{GPE}) on the ladder and crossed diagrams was studied in \cite{Wellens08, Wellens09a, Wellens09b} in the stationary limit where $\psi(\bi{r},t)=\psi(\bi{r})e^{i\epsilon t}$. It was demonstrated that the leading-order nonlinear correction to the series of ladder diagrams \textit{vanishes}. In other words, in a first approximation diffusive motion is \textit{not affected} by the nonlinearity. For a wave packet, this conclusion remains true provided the wave packet remains quasi-monochromatic, i.e. as long as its energy distribution does not significantly change in time. We will make this assumption from here on, posponing the discussion of its validity to Sec. \ref{Validity_Sec}. In the diffusive limit (interference terms neglected), the density is then again approximately given by
\begin{equation}
\label{n_convolv}
n(\bi{r},t)\simeq
\int d^d\bi{r}'\,
P(\bi{r}',\bi{r},t)
|\phi(\bi{r}')|^2,
\end{equation}
where the diffuson $P$ obeys Eq. (\ref{diff_eq}) in the hydrodynamic regime. 

\begin{figure}
\centering
\includegraphics[width=12cm]{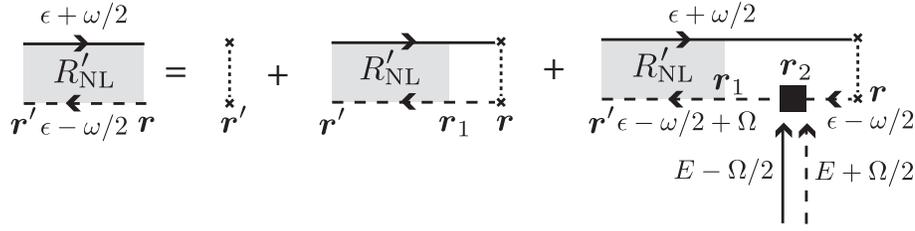}
\caption{\label{NL_BS_fig} 
Iterative equation obeyed by the kernel $R'_\mathrm{NL}$. The black square together with its two incoming arrows symbolize the nonlinear potential $g n_E(\bi{r}_2,\Omega)$.
}  
\end{figure}

Let us now discuss the Cooperon. Its dependence on the nonlinearity was addressed in a number of works dealing with nonlinear coherent backscattering \cite{Agranovich91, Hartung08, Wellens08, Wellens09a, Wellens09b, Hartmann12}, with the following  conclusions. Due to the term $g|\psi|^2$, the nonlinear Born series constructed from Eq. (\ref{GPE}) generates diagrams which combine more than two (advanced and retarded) Green's functions after a disorder average. Consequently, when $g\ne0$ weak localization is strictly speaking no longer a two-wave interference involving a scattering path and its time-reversed partner, but rather a multiple-wave process involving sequences of crossed diagrams connected by scattering events from the nonlinear potential $g|\psi|^2$ in many possible ways. The task of accounting  for all these combinations was accomplished in \cite{Hartung08, Wellens08, Wellens09a, Wellens09b, Hartmann12}. The net result leads to a simple form for the nonlinear version of the Cooperon:
\begin{equation}
\label{Fdef}
P'(\bi{r}',\bi{r},t)=(\tau/\gamma)\Re\left[R'_\mathrm{NL}(\bi{r}',\bi{r},t)\right],
\end{equation}
where the kernel $R'_\mathrm{NL}$ obeys the iterative equation shown diagrammatically in Fig. \ref{NL_BS_fig} and expressed as
\begin{eqnarray}
\label{NL_BSE_eq}
 \fl R'_\mathrm{NL}(\bi{r}',\bi{r},\omega)&=&\gamma\delta(\bi{r}-\bi{r}')+
\gamma\int d^d\bi{r}_1 \overline{G}^R_{\epsilon+\omega/2}(\bi{r}_1,\bi{r})\overline{G}^A_{\epsilon-\omega/2}(\bi{r}_1,\bi{r}) R'_\mathrm{NL}(\bi{r}',\bi{r}_1,\omega)+\nonumber\\
&&  g\gamma\int d^d\bi{r}_1  d^d\bi{r}_2
 \frac{d\Omega}{2\pi} \frac{dE}{2\pi}
\overline{G}^R_{\epsilon+\omega/2}(\bi{r}_1,\bi{r})
\overline{G}^A_{\epsilon-\omega/2+\Omega}(\bi{r}_1,\bi{r}_2)
\overline{G}^A_{\epsilon-\omega/2}(\bi{r}_2,\bi{r})\times\nonumber\\ 
&& R'_\mathrm{NL}(\bi{r}',\bi{r}_1,\omega-\Omega)n_E(\bi{r}_2,\Omega),
\end{eqnarray}
where $n_E(\bi{r}_2,\Omega)\equiv\overline{\psi_{E+\Omega/2}(\bi{r}_2)\psi^*_{E-\Omega/2}(\bi{r}_2)}$, with $\psi_\epsilon(\bi{r})$ defined by Eq. (\ref{psi_G}). 
The third term in the right-hand side of Eq. (\ref{NL_BSE_eq}) accounts for the possibility of scattering from the nonlinear potential in one of the two counter-propagating paths, and Eq. (\ref{Fdef}) guarantees that the Cooperon $P'$ is real\footnote{From a technical point of view, this real part originates from the removal of certain classes of diagrams involving closed loops and not generated by the Born series. This imposes to define an auxillary Bethe-Salpeter equation that cancels the imaginary part of $R'_\mathrm{NL}(\bi{r}',\bi{r},t)$ (for the details of this procedure, see for instance \cite{Wellens09b}).}.

To proceed further, we drop the $\Omega$ dependence in the Green's function $\overline{G}^A_{\epsilon-\omega/2+\Omega}$ in Eq. (\ref{NL_BSE_eq}). In turn, this amounts to neglecting the mixing of energy components during the time evolution. We also expand $n_E(\bi{r}_2,\Omega)$ around $\bi{r}_2\simeq \bi{r}_1\simeq \bi{r}$ (hydrodynamic approximation). This allows us to perform the integrals over $\bi{r}_2$ and $\Omega$ to obtain
\begin{eqnarray}
\label{NL_BSE_simple}
\fl && R'_\mathrm{NL}(\bi{r}',\bi{r},\omega)\simeq\gamma\delta(\bi{r}-\bi{r}')+\nonumber\\
&&\gamma\int d^d\bi{r}_1 \overline{G}_{\epsilon+\omega/2}(\bi{r}_1,\bi{r})\overline{G}^A_{\epsilon-\omega/2}(\bi{r}_1,\bi{r}) 
\left[1+i\tau gn(\bi{r}_1,\omega)*\right]
R'_\mathrm{NL}(\bi{r}',\bi{r}_1,\omega),\\\nonumber
\end{eqnarray}
where $n(\bi{r}_1,\omega)*R'_\mathrm{NL}(\bi{r}',\bi{r}_1,\omega)\equiv \int d\Omega/(2\pi)\,n(\bi{r}_1,\Omega)R'_\mathrm{NL}(\bi{r}',\bi{r}_1,\omega-\Omega)$ and $n(\bi{r}_1,\Omega)\equiv\int dE/(2\pi)n_E(\bi{r}_1,\Omega)$.
Noticing that $1+i\tau gn\simeq e^{i\phi}$ with $\phi=\tau g n$, we infer that the nonlinearity plays the role of a (density-dependent) \textit{dephasing} \cite{Wellens09b} \footnote{This also justifies a posteriori why, to leading order, the nonlinearity does not alter the diffuson.}. The latter is responsible for the decay of the coherent backscattering effet observed numerically \cite{Hartung08, Wellens08, Wellens09a, Wellens09b}. We will see below that it is also the primary mechanism by which Anderson localization is altered by the nonlinearity.
Note that the nonlinear term in Eq. (\ref{NL_BSE_simple}) manifests itself via a convolution operator $*$ in frequency space, which is reminiscent of the multiplicative nonlinear potential in Eq. (\ref{GPE}). Using again the hydrodynamic approximation, we finally expand $R'_\mathrm{NL}(\bi{r}',\bi{r}_1,\omega)$ around $\bi{r}_1\simeq \bi{r}$ as well as the Green's functions for $\omega\tau\ll1$. This yields
\begin{equation}
\label{PC_eq}
\left[-i\omega-D_0\boldsymbol\nabla^2_\bi{r}-i g n(\bi{r},\omega)*\right]R'_\mathrm{NL}(\bi{r}',\bi{r},\omega)=(\gamma/\tau)\delta(\bi{r}-\bi{r}').
\end{equation}
Eq. (\ref{PC_eq}) will be our basic ingredient to construct a nonlinear SCT of localization.

\subsection{Nonlinear self-consistent theory}

A first idea to generalize the SCT to the nonlinear regime would be to merely write that $P(\bi{q},\omega)=1/[-i\omega+D(\omega)\bi{q}^2]$, with $1/D(\omega)=1/D_0+1/(\pi\rho D_0)\int d^d\bi{Q}/(2\pi)^dP'(\bi{Q},\omega)$ and $P'(\bi{Q},\omega)$ solution of Eqs. (\ref{Fdef}) and (\ref{PC_eq}). However, such a prescription is not rigorous for the following reason.
In the self-consistent scheme depicted in the right panel of Fig. \ref{Loops_on_loops}, localization loops are nested according to a diffuson/Cooperon alternation. Since in the nonlinear regime these two objects are different, it is necessary to define \textit{two} (different) diffusion coefficients, $D$ for the diffuson and $D'$ for the Cooperon. This idea was originally suggested in \cite{Eckert10}. In spirit, it is similar to what was proposed in \cite{Yoshioka81, Ono81} to account for magnetic-fields effects in the SCT. 
It leads us to introduce a \emph{set} of self-consistent equations:
\begin{subnumcases}{}
  \left[-i\omega- D(\omega)\boldsymbol\nabla^2_{\bi{r}}\right]P(\bi{r}',\bi{r},\omega)=\delta(\bi{r}-\bi{r}') \label{NL_SCEa}\\
        \frac{1}{D(\omega)}=\frac{1}{D_0}+\frac{1}{\pi\rho D_0}P'(\bi{r},\bi{r},\omega) \label{NL_SCEb}\\
  \left[-i\omega-D'(\omega)\boldsymbol\nabla^2_{\bi{r}}-i g n(\bi{r},\omega)*\right]R'_\mathrm{NL}(\bi{r}',\bi{r},\omega)=(\gamma/\tau)\delta(\bi{r}-\bi{r}') \label{NL_SCEc}\\
        \frac{1}{D'(\omega)}=\frac{1}{D_0}+\frac{1}{\pi\rho D_0}P(\bi{r},\bi{r},\omega), \label{NL_SCEd}
\end{subnumcases}
where $P'(\bi{r}',\bi{r},t)=(\tau/\gamma)\Re\left[R'_\mathrm{NL}(\bi{r}',\bi{r},t)\right]$ and $n(\bi{r},t)=\int d^d\bi{r}''P(\bi{r}'',\bi{r},t)|\phi(\bi{r}'')|^2$, with an initial wave packet $\phi(\bi{r}'')$ centered at $\bi{r}''=\bi{r}'$. 
Eqs. (\ref{NL_SCEa}--\ref{NL_SCEd}) constitute a generalization of the SCT to the nonlinear regime. 
One easily verifies that Eqs. (\ref{SCTa}) and (\ref{SCTb}) are recovered when $g=0$ (in which case $D=D'$).

\section{Approximate solution}
\label{Approx_sol_Sec}

\subsection{Ansatz}
\label{Ansatz_Sec}

To unveil the predictions of the nonlinear SCT for the dynamics of wave packets, it is in principle required to solve the set of equations (\ref{NL_SCEa}-\ref{NL_SCEd}) with the initial condition $|\phi(\bi{r}'')|^2=N \delta(\bi{r}''-\bi{r}')$. Due to the spatial dependence of the nonlinear term in Eq. (\ref{NL_SCEc}) however, this seems to be a hard task. To nevertheless extract the physical content of these equations, we propose an approximate resolution method: we first solve the SCT for an initial wave packet $|\phi(\bi{r})|^2\sim N/L^d$ of very large size $L\to\infty$, so that $g n(\bi{r},\omega)*R'_\mathrm{NL}(\bi{r}',\bi{r},\omega)=(N/L^d) R'_\mathrm{NL}(\bi{r}',\bi{r},\omega)$, and then infer the results for an initially narrow wave packet by substituting the spatial extent  $L_\omega=\sqrt{D(\omega)/(-i\omega)}$ of the propagator $P(\bi{r}', \bi{r},\omega)$ for $L$. This Ansatz allows us to rewrite Eqs. (\ref{NL_SCEa}-\ref{NL_SCEd}) in Fourier space as:
\begin{subnumcases}{}
  \left[-i\omega+ D(\omega)\bi{q}^2\right]P(\bi{q},\omega)=1 \label{NL_SCE_n0a} \\
  \frac{1}{D(\omega)}=\frac{1}{D_0}+\frac{1}{\pi\rho D_0}\displaystyle\int\frac{d^d\bi{Q}}{(2\pi)^d}P'(\bi{Q},\omega) \label{NL_SCE_n0b} \\
  \left[-i\omega+ D'(\omega)\bi{q}^2-\frac{ig N}{L_\omega^d}\right]R'_\mathrm{NL}(\bi{q},\omega)=\gamma/\tau  \label{NL_SCE_n0c}\\
        \frac{1}{D'(\omega)}=\frac{1}{D_0}+\frac{1}{\pi\rho D_0}\displaystyle\int\frac{d^d\bi{Q}}{(2\pi)^d}P(\bi{Q},\omega).  \label{NL_SCE_n0d}
\end{subnumcases}
Using that $P'(\bi{q},\omega)=\tau/(2\gamma)\left[R'_\mathrm{NL}(\bi{q},\omega)+R^{\prime*}_\mathrm{NL}(\bi{q},-\omega)\right]$ [see Eq. (\ref{Fdef})] and $D'(-\omega)^*=D'(\omega)$, we infer from Eq. (\ref{NL_SCE_n0c}):
\begin{equation}
\label{eqPc}
P'(\bi{q},\omega)=\frac{-i\omega+D'(\omega)\bi{q}^2}{[-i\omega+D'(\omega)\bi{q}^2]^2+[g N/L_\omega^d]^2}.
\end{equation}
At this stage, an important comment is in order: since $P'(\bi{q},\omega)$ only depends on $g^2$, the sign of the nonlinearity plays \textit{no role} at our level of approximation. This is consistent with the picture of a weak nonlinearity acting as a dephasing. For this reason, from here on we will assume $g>0$ without loss of generality. 

\subsection{Results for $D(\omega)$ in dimension 3}
\label{3D_results_Sec}

We now discuss the solution of Eqs (\ref{NL_SCE_n0a}-\ref{NL_SCE_n0d}) for $D(\omega)$ in 3D. Integrals over momenta in Eqs. (\ref{NL_SCE_n0b}) and (\ref{NL_SCE_n0d}) need to be regularized by an ultraviolet cutoff $Q_\mathrm{max}$. This cutoff has the same origin as in the usual SCT: the breakdown of the hydrodynamic approximation at short scales (see Sec. \ref{SCT_predictions}). 
Since any alteration of the mean free path by the nonlinearity is a second-order effect compared to the dephasing-like mechanism discussed in this paper  \cite{Spivak00, Cherroret09}, it is reasonable to keep the same cutoff  as for $g=0$: $Q_\mathrm{max}=\pi/(3\ell)$.

\begin{figure}[h]
\centering
\includegraphics[width=9.cm]{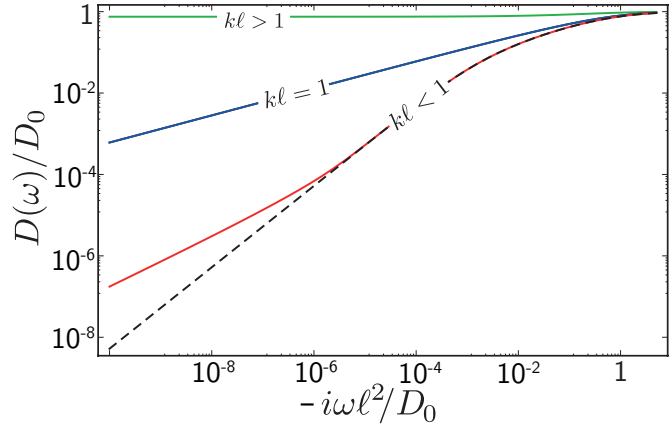}
\caption{\label{Nonlinear_D_3DLoc} 
Solution $D(\omega)$ of Eqs. (\ref{NL_SCE_n0a}--\ref{NL_SCE_n0d}) as a function of $-i\omega\ell^2/D_0$ for $kl=0.9$, $kl=1$ and $k\ell=2$ in 3D (solid curves).  The value of the nonlinearity is set to $gN/(D_0\ell)=10^{-3}$. For $k\ell>1$ and $k\ell=1$, there is no visible difference with the case $g=0$ (curves for $g=0$ and $g\ne 0$ nearly superimpose). In contrast, when $k\ell<1$ the linear (dashed curve) and nonlinear predictions clearly deviate at small frequencies.
}  
\end{figure}
Substitution of Eq. (\ref{eqPc}) for $P'$ in Eq. (\ref{NL_SCE_n0b}) yields a relation between $D(\omega)$ and $D'(\omega)$, which allows us to eliminate $D'(\omega)$ in Eq. (\ref{NL_SCE_n0d}). We thus obtain an implicit equation for $D(\omega)$ only, which we can solve. The result is shown in Fig. \ref{Nonlinear_D_3DLoc} as a function of $-i\omega\ell^2/D_0$ for $k\ell>1$, $k\ell=1$ and $k\ell<1$. As compared to the linear case (Fig. \ref{Linear_SCT}), $D(\omega)$ remains qualitatively not affected by the nonlinearity for $k\ell>1$ and $k\ell=1$. This result can be interpreted by the fact that, in these regimes, the dephasing $\phi\propto g n(\bi{r},t)$ quickly decays to zero as the wave packet spreads (dilution of the nonlinearity). In contrast, for $k\ell<1$ we observe a marked deviation from the linear prediction at a characteristic time scale $\tau_\mathrm{NL}$. Leaving out the numerical prefactors, we find:
\begin{equation}
\label{tauNL_def}
\tau_\mathrm{NL}=\frac{\xi^3}{gN}.
\end{equation}
In energetic terms, $\tau_\mathrm{NL}^{-1}$ is the typical interaction energy in the localization volume $\xi^3$. For $\omega\tau_\mathrm{NL}\ll1$, the asymptotic law $D(\omega)=-i\omega\xi^2$ is changed to
\begin{equation}
\label{Asymptotics_DL_3D_NWP}
D(\omega)\simeq\left(\frac{gN\xi^2}{8}\right)^{2/5}(-i\omega)^{3/5}.
\end{equation}
According to Eq. (\ref{D_width}), this law describes a wave packet whose mean-square width scales as $\langle\bi{r}^2(t)\rangle\propto t^{2/5}$. In other words, at long times Anderson localization is destroyed by the nonlinearity and replaced by algebraic \textit{subdiffusion}. This phenomenon is the result of a trade-off between interference due to disorder and dephasing due to interactions as the wave packet spreads: on the one hand, interference tends to localize the packet and thus to reinforce the dephasing mechanism by preventing $\phi$ from decreasing to zero as time grows. On the other hand, interactions tend to delocalize the packet, which makes $\phi$ decrease and in turn reinforces interference. Note that there is no threshold for $g$ within our formalism: as soon as $g$ is nonzero, the system will eventually end up in a subdiffusive regime when $k\ell<1$. Since the establishment of subdiffusion occurs when $t>\tau_\mathrm{NL}\sim 1/g$ though, it may never be seen if $g$ is too small.

As outlined above, the nonlinearity has almost no effect on the diffusion coefficient for $k\ell>1$ and $k\ell=1$ (see Fig. \ref{Nonlinear_D_3DLoc}). We can quantify this statement by solving Eqs. (\ref{NL_SCE_n0a}--\ref{NL_SCE_n0d}) analytically at low frequencies. For $k\ell\gg1$ we find
\begin{equation}
\label{Asymptotics_kllarge}
\fl D(\omega\rightarrow 0)\simeq
1-\frac{1}{(k\ell)^2}
\left[1-\frac{3}{2}\left(\frac{-i\omega\ell^2}{D_0}\right)^{1/2}
	-\frac{3}{16}\left(\frac{gN}{D_0\ell}\right)^2\left(\frac{-i\omega\ell^2}{D_0}\right)^{3/2}
\right].
\end{equation}
The term proportional to $1/(k\ell)^2$ is the weak localization correction. 
It is indeed very weakly modified by the nonlinearity, and moreover the nonlinear correction becomes weaker as time grows beyond $\tau$. Exactly at $k\ell =1$ finally, we find
\begin{equation}
\label{Asymptotics_kl1}
D(\omega\rightarrow 0)\simeq
\left(\frac{3 D_0\ell}{2}\right)^{2/3}
\left[1-\frac{1}{54}\left(\frac{gN}{D_0\ell}\right)^2
\right](-i\omega)^{1/3}.
\end{equation}
Thus, the nonlinearity slightly decreases the constant prefactor in the diffusion coefficient, but does not modify the characteristic $\omega^{1/3}$ dependence and thus the critical law $\langle\bi{r}^2(t)\rangle\propto t^{2/3}$. 
\begin{figure}[h]
\centering
\includegraphics[width=7.5cm]{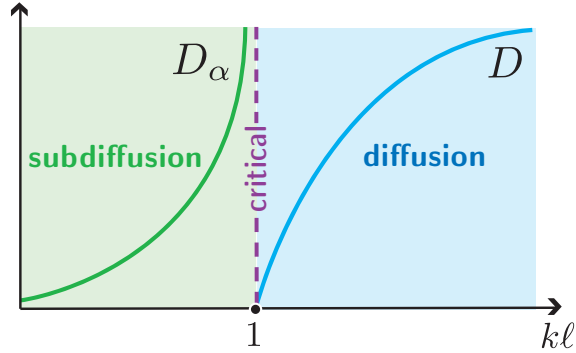}
\caption{\label{Phase_diagram2} 
Illustration of the subdiffusion-diffusion transition that replaces the Anderson transition for $g\ne 0$. The transition is characterized by a vanishing of the diffusion coefficient $D$ as $k\ell\to 1^+$ (as in the linear case) and by a divergence of the subdiffusion coefficient $D_\alpha$ as $k\ell\to 1^-$.
}  
\end{figure}

\subsection{A subdiffusion-diffusion transition}

Eq. (\ref{Asymptotics_kllarge}) additionally shows that, at fixed $g\ne0$, $D(\omega\rightarrow0)\equiv D\to 0$ at $k\ell=1$. This signals the persistence of a phase transition at $k\ell=1$ when $g\ne0$. However, this transition is fundamentally different from the Anderson transition since it now separates a regime of diffusion from a regime of subdiffusion. Before investigating its critical manifestations, let us first recall the essential properties of the Anderson transition that occurs when $g=0$. For a spreading wave packet, this transition shows up at  $t\to\infty$~\cite{Shapiro82} and is characterized by two critical exponents $s$ and $\nu$. $s$ controls the vanishing of the diffusion coefficient as the critical point is approached from above ($k\ell\to 1^+$), according to
\begin{equation}
\frac{\langle\bi{r}^{2}(t)\rangle}{t}\underset{t\to\infty}{\sim}D\propto(k\ell-1)^{s},\label{r2_diff}
\end{equation}
while $\nu$ controls the divergence of the localization length as the critical point is approached from below ($k\ell\to 1^-$):
\begin{equation}
\sqrt{\langle\bi{r}^{2}(t)\rangle}\underset{t\to\infty}{\sim}\xi\propto\frac{1}{(1-k\ell)^{\nu}}.
\label{r2_loc}
\end{equation}
Although the SCT is not able to capture the correct value of $s$ and $\nu$ (it gives $s=\nu=1$, whereas $s=\nu\simeq1.58$ according to numerical simulations \cite{Slevin10}), it predicts well the vanishing of $D$ and the divergence of $\xi$ at $k\ell=1$, as well as the Wegner law \cite{Vollhardt82} which states that $s$ and $\nu$ are equal in 3D~\cite{Wegner76}. We can take advantage of this to make some predictions on the critical properties of the subdiffusion-diffusion transition that replaces the Anderson transition when $g\ne 0$. Indeed, from Eqs. (\ref{Asymptotics_DL_3D_NWP}) and (\ref{D_width}), we find $\langle\bi{r}^2(t)\rangle/t^\alpha\sim g^\alpha\xi^{2\alpha}$ on the subdiffusive side, where $\alpha=2/5$ is the subdiffusion coefficient. Using Eq. (\ref{r2_loc}) for $\xi$, we then infer
\begin{equation}
D_{\alpha}\equiv\frac{\langle\bi{r}^{2}(t)\rangle}{t^{\alpha}}\underset{t\to\infty}{\propto}\frac{g^{\alpha}}{(1-k\ell)^{2\alpha\nu}},\ (g\ne0,\, k\ell\to 1^-).
\label{r2_nonlinear}
\end{equation}
Thus, when $g\ne0$ the \emph{subdiffusion coefficient} $D_\alpha$ diverges at $k\ell=1$, with a critical exponent $2\alpha\nu$. Since on the diffusive side Eq. (\ref{r2_diff}) is unchanged, the natural generalization of the Wegner law to the case $g\ne 0$ reads:
\begin{equation}
s=2\alpha\nu.
\label{r2_nonlinear}
\end{equation}
The subdiffusion-diffusion transition is illustrated in Fig. \ref{Phase_diagram2}. 

\section{Two-parameter scaling theory}
\label{TPST}

An interesting question is how the one-parameter scaling theory of localization \cite{Abrahams79} is modified by the nonlinearity. In its historical version, the scaling theory describes transport in a disordered conductor of size $L$  in terms of a single parameter, the dimensionless Thouless conductance $G$ \footnote{We use the notation $G$ instead of the more usual $g$, to avoid any confusion with the nonlinear parameter.} \cite{Thouless74}. 
\begin{figure}[h]
\centering
\includegraphics[width=10cm]{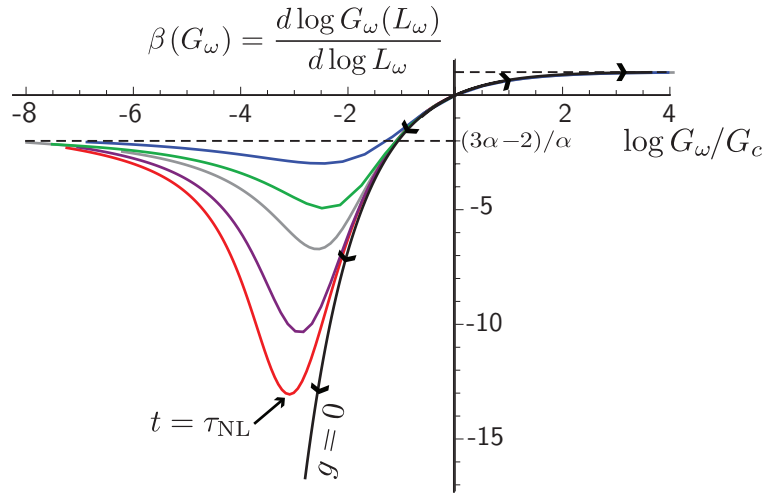}
\caption{\label{beta} 
Scaling function (\ref{betaLomega}), derived from Eqs. (\ref{NL_SCE_n0a}--\ref{NL_SCE_n0d}) in 3D. For $g=0$ (black curve), $\beta$ is monotonic. It vanishes at $G_c\sim 1$, is positive for $G_\omega> G_c$ and negative for $G_\omega<G_c$. Colored curves correspond to nonzero values of the parameter $\lambda=gN/(D_0\ell)$ ($\lambda=0.035$, 0.05, 0.1, 0.17 and 1 from bottom to top). The nonlinearity gives birth to a minimum (associated with the characteristic time $\tau_\mathrm{NL}$) and to a breakdown of monotonicity. For $G_\omega\to 0$, $\beta\to(3\alpha-2)/\alpha$ and the system is in the subdiffusive phase.
}  
\end{figure}
We can extend this formalism to the time-dependent scenario we are concerned with in the following way: we take $L_\omega=\sqrt{D(\omega)/(-i\omega)}$ (the wave-packet width in frequency space) for the system size, and we define a dimensionless ``conductance'' as $G_\omega=L_\omega^{d-2}D(\omega)$, by analogy with a true conductor \cite{Vollhardt82, Ostrovsky13}. According to the one-parameter scaling hypothesis, when $g=0$ $L_\omega$ and $G_\omega$ are related by the renormalization-group equation 
 \begin{equation}
 \label{betaLomega}
\frac{d\log G_\omega(L_\omega)}{d \log L_\omega}=\beta(G_\omega),
\end{equation}
where $G_\omega$ depends only on $L_\omega$. The function $\beta(G_\omega)$ can be derived from Eq. (\ref{SCTb}) \cite{Vollhardt82}. The result is shown in Fig. \ref{beta} as a black curve. We find that by changing independently $k\ell$ and $\omega$ one always moves along a single curve, which confirms the one-parameter scaling hypothesis. As in a conductor, we also have
$\beta(G_\omega)\to 1$ far in the diffusive phase (defined by $\beta>0$) and  $\beta(G_\omega)\to -\infty$ far in the localized phase (defined by $\beta<0$)
. The scaling function changes of sign at $G_\omega=G_c\sim 1$, the unstable fixed point of the Anderson transition \cite{Abrahams79}. The arrows in Fig. \ref{beta} indicate the direction of the flow as $L_\omega$ is increased.

In Fig. \ref{beta} we also show $\beta(G_\omega)$ for $g\ne0$, obtained by solving Eqs. (\ref{NL_SCE_n0a}--\ref{NL_SCE_n0d}) in 3D (colored curves). In line with the results of Sec. \ref{3D_results_Sec}, on the diffusive side of the transition the nonlinearity does not qualitatively modify the scaling function as well as the location of the critical point. On the other side in contrast, the nonlinearity gives birth to a minimum and leads to a breakdown of monotonicity. This minimum corresponds to the characteristic time $t\sim \tau_\mathrm{NL}=\xi^3/(gN)$ associated with the crossover of Fig. \ref{Nonlinear_D_3DLoc}, which separates a regime of transient localization and the asymptotic subdiffusive limit where $\beta(G_\omega)\to (3\alpha-2)/\alpha$.
Fig. \ref{beta} also shows that different values of $g$ generate different scaling functions, which suggests that a single-parameter description of the system is no longer possible in the presence of nonlinearity. Inspection of Eqs. (\ref{NL_SCE_n0a}--\ref{NL_SCE_n0d}) reveals that nonlinear corrections systematically  appear as terms proportional to $\Gamma_\omega=\lambda/G_\omega$, where $\lambda=gN/(D_0\ell)$. It is therefore natural to define this quantity as the additional scaling parameter of the  nonlinear problem, yielding the \emph{two-parameter scaling theory}
\begin{subnumcases}{}
 \frac{d\log G_\omega}{d \log L_\omega}=\beta_1(G_\omega, \Gamma_\omega) \label{twop_scaling1} \\
\frac{d\log \Gamma_\omega}{d \log L_\omega}=\beta_2(G_\omega, \Gamma_\omega). \label{twop_scaling2}
 \end{subnumcases}
Eqs. (\ref{twop_scaling1}) and (\ref{twop_scaling2}) generalize Eq. (\ref{betaLomega}) to $g\ne0$. They introduce two scaling functions $\beta_1$ and $\beta_2$ which are in fact trivially related due to the relation $\Gamma_\omega=\lambda/G_\omega$: $\beta_1=-\beta_2\equiv \beta$.
\begin{figure}[h]
\centering
\includegraphics[width=9cm]{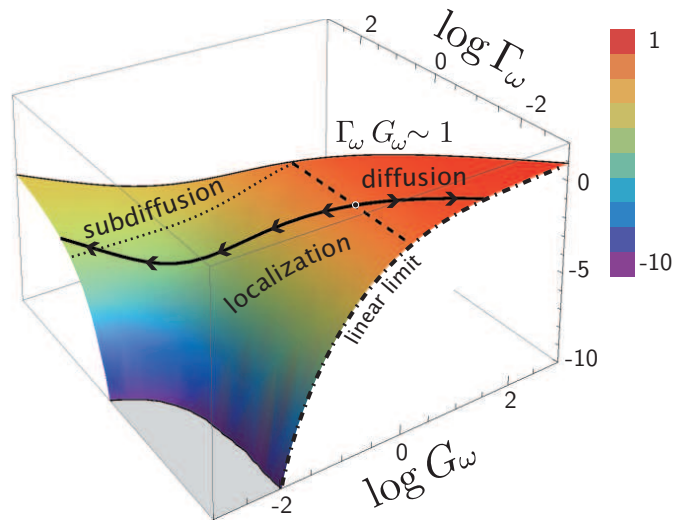}
\caption{\label{Phase_diagram} 
Two-parameter scaling function $\beta(G_\omega,{\Gamma_\omega})$ derived from Eqs. (\ref{NL_SCE_n0a}--\ref{NL_SCE_n0d}) in 3D.
The solid curve is the trace of the $\beta$ curves of Fig.~\ref{beta} that correspond to fixed values of $\lambda=gN/(D_{0}\ell)$. The dashed line gives the position of the critical point $G_\omega=G_c\simeq1$ (where $\beta=0$), and the dotted curve indicates the crossover $\Gamma_\omega\simeq1$ between localization and subdiffusion. The dotted-dashed curve shows the scaling function in the limit $g\to0$. Beyond the border $\Gamma_\omega G_\omega\equiv\lambda\sim1$, the nonlinearity can no longer be considered weak.}  
\end{figure}
A 3D plot of $\beta(G_\omega,\Gamma_\omega)$ is displayed in Fig. \ref{Phase_diagram}. The three regions of classical diffusion, Anderson localization and nonlinear subdiffusion are highlighted. The solid curve is the trace of the $\beta$ curves of Fig. \ref{beta} that correspond to fixed values of $\lambda$, i.e. to $\Gamma_\omega G_\omega=$ constant. The dashed line parallel to the $\Gamma_\omega$ axis indicates the position of the critical point $G_\omega=G_c\simeq 1$ where $\beta$ vanishes. The dotted curve shows the boundary $\Gamma_\omega\simeq 1$. This boundary approximately coincides with the minimum of the curves in Fig. \ref{beta}. Remember, however, that the line $\Gamma_\omega\simeq 1$ is not associated with a phase transition ($\beta\neq 0$) but with the crossover between localization and subdiffusion that occurs at $t\sim\tau_\mathrm{NL}$. In the region $\lambda=G_\omega\Gamma_\omega>1$, the nonlinearity can no longer be considered weak.

\section{Accuracy and limitations of the theory}
\label{Validity_Sec}

To summarize, for wave packets spreading in random potentials in 3D in the presence of a quadratic nonlinear potential, our nonlinear SCT predicts:
 (i) a robustness of diffusive (at $k\ell>1$) and subdiffusive motion (at $k\ell=1$)  (ii) a robustness of the position of the critical point (iii) a destruction of localization in favor of subdiffusion for $t>\tau_\mathrm{NL}\propto 1/g$ when $k\ell<1$ (iv) the divergence of the subdiffusion coefficient at $k\ell=1$ and (v) a breakdown of the one-parameter scaling description. 

From a numerical point of view, simulations of wave-packet spreading have been carried out in the quasi-periodic kicked rotor with three incommensurate frequencies in  \cite{Cherroret14}. In the absence of nonlinearity, this model is formally equivalent to a spatially disordered system in 3D, except that localization takes place in momentum space rather than in configuration space. After turning on a quadratic nonlinearity in this system, the properties (i), (ii), (iii) and (v) and, in particular, the breakdown of monotonicity of the scaling function  (Fig. \ref{beta}) have been confirmed. Near the critical point, signatures of an algebraic scaling of the subdiffusion coefficient have also been found. 

In a spatially disordered system, it should be noted that because of the phenomenon of thermalization of the particle energy distribution occuring as time grows (neglected in this work), the nonlinearity-driven subdiffusion-diffusion transition may be eventually smoothed out at long times. For a weak enough nonlinearity though, the time scale for thermalization is $\sim 1/g^2\gg1/g\sim \tau_\mathrm{NL}$  (see also below) \cite{Cherroret15}, so that the nonlinear transition should be visible within an intermediate time window.

\begin{figure}[h]
\centering
\includegraphics[width=8cm]{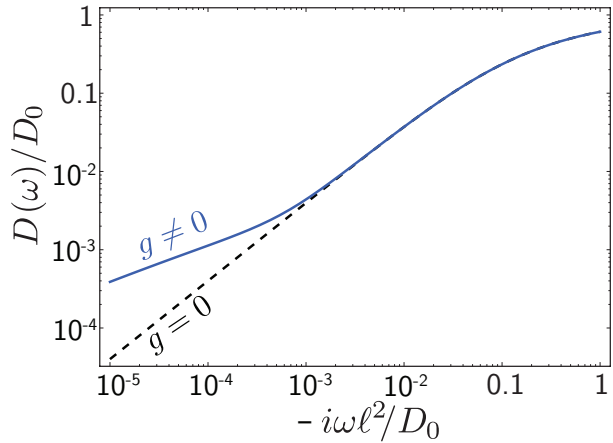}
\caption{\label{Nonlinear_D_1D} 
Solution $D(\omega)$ of Eqs. (\ref{NL_SCE_n0a}--\ref{NL_SCE_n0d}) as a function of $-i\omega\ell^2/D_0$ in 1D (solid curve).
The dashed curve is the linear prediction, which displays the low-frequency asymptotic law $D(\omega) = -i \omega\xi^2$. The value of the nonlinearity is set to $gN\ell/D_0 = 10^{-3}$.}  
\end{figure}

To further assess the accuracy of the nonlinear SCT, it is also instructive to discuss the one-dimensional case. The solution of Eqs. (\ref{NL_SCE_n0a}--\ref{NL_SCE_n0d}) for $D(\omega)$ in 1D is shown in Fig. \ref{Nonlinear_D_1D}.
Again, Anderson localization is destroyed and replaced by subdiffusion at long times $t\gg\tau_\mathrm{NL}^\mathrm{1D}=\xi/(gN)$. Indeed, for $\omega \tau_\mathrm{NL}^\mathrm{1D}\ll 1$ we find $D(\omega)=(2gN\xi^3)^{1/2}(-i\omega)^{1/2}$, corresponding to $\langle\bi{r}^2(t)\rangle\sim t^{1/2}$. The destruction of localization in favor of subdiffusion in 1D was observed in numerical simulations of the discrete nonlinear Schr\"odinger equation \cite{Kopidakis08, Pikovsky08}, and a similar time scale $\sim1/g$ beyond which subdiffusion prevails was predicted in \cite{Iomin09, Iomin10} by means of a probabilistic approach. While the precise value of the subdiffusive exponent $\alpha$ is still debated, a value between 0.3 and 0.4 was identified \cite{Kopidakis08, Pikovsky08}. Compared to this estimation, the prediction 1/2 of our approach is thus slightly too large, and the same conclusion might be true in 2D and 3D. Several approximations used in this paper may explain this overestimation. The main one is the limit of ``quasi-monochromatic'' wave packet: we have assumed that the particle energy distribution is very narrow (Sec. \ref{MonoC}), and that it remains so at any time (Sec. \ref{LCD_nonlinear_sec}). Such an approximation breaks down at very long times. For instance, for an initial plane wave evolving in momentum space it was found that the energy distribution thermalizes over a time scale $\sim 1/g^2$ \cite{Cherroret15}. How the change of the energy distribution affects the value of $\alpha$ in our case remains to ascertain. The approximate resolution method we have used to solve the nonlinear equations (Sec. \ref{Ansatz_Sec}) can be another reason for the too large subdiffusive exponent. Indeed, in this rather crude approach we have neglected both the $\bi{r}$ dependence and the nonlocal character of the frequency dependence of the nonlinear term in Eq. (\ref{NL_SCEc}), which may lead to an overestimation of the weight of the nonlinearity.

\section{Conclusion}

We have developed a self-consistent theory of Anderson localization for a particle evolving in a random potential  according to the Gross-Pitaevskii equation. For wave packets spreading in 3D, the theory predicts a robustness of diffusive motion above the critical point and a robustness of the location of the mobility edge. In strong contrast, below the critical point Anderson localization is destroyed and replaced by subdiffusion. The Anderson transition is thus changed to a subdiffusion-diffusion transition, associated with the emergence of novel critical properties and to a breakdown of the one-parameter scaling description. In 1D, our approach also predicts the emergence of subdiffusive motion at long times with, however, a slightly too large subdiffusive exponent. This overestimation might be due to the approximations on which the theory is based on, and especially the stationarity and narrowness of the particle energy distribution in the course of time. An extension of the theory relaxing this assumption is an interesting challenge for the future. In the diffusive regime, first steps in this direction have been already accomplished \cite{Schwiete10, Cherroret11, Schwiete13}.

\ack

The author is indebted to Thomas Wellens for his explanations of the nonlinear diagrammatic approach to disordered media prior the beginning of this work. Many discussions with Dominique Delande, Jean Claude Garreau, Beno\^it Vermersch and Tony Prat are gratefully acknowledged.

\section*{References}


\end{document}